\documentclass[aps,prb,superscriptaddress,twocolumn,final,floatfix,showpacs]{revtex4}
\usepackage{amsmath}
\usepackage{graphicx}
\usepackage{longtable}
\usepackage{latexsym}

\begin{document}

\title{First-principles calculation of the band offset at BaO/BaTiO$_3$
           and SrO/SrTiO$_3$ interfaces.}

\author{ Javier Junquera }
\affiliation{ D\'epartement de Physique, Universit\'e de Li\`ege,
                  B\^atiment B-5, B-4000 Sart-Tilman, Belgium }

\author{ Magali Zimmer }
\affiliation{ D\'epartement de Physique, Universit\'e de Li\`ege,
                  B\^atiment B-5, B-4000 Sart-Tilman, Belgium }

\author{ Pablo Ordej\'on }
\affiliation{ Institut de Ci\`encia de Materials de Barcelona, CSIC,
                  Campus de la UAB, Bellaterra,
                  08193 Barcelona, Spain }

\author{ Philippe Ghosez }
\affiliation{ D\'epartement de Physique, Universit\'e de Li\`ege,
                  B\^atiment B-5, B-4000 Sart-Tilman, Belgium }

\date{\today}

\begin{abstract}
     We report first-principles density-functional pseudopotential
     calculations on
     the atomic structures, electronic properties, and band offsets of
     BaO/BaTiO$_3$ and SrO/SrTiO$_3$ nanosized 
     heterojunctions grown on top of a silicon substrate.
     The density of states at the junction does not reveal any
     electronic induced interface states. 
     A dominant perovskite character is found at the interface layer.
     The tunability of the band offset with the strain conditions
     imposed by the substrate is studied.
     Using previously reported theoretical data available for Si/SrO,
     Si/BaO and BaTiO$_{3}$/SrRuO$_{3}$ interfaces
     we extrapolate a value for the band alignments along the whole
     gate stacks of technological interest:
     Si/SrO/SrTiO$_3$ and Si/BaO/BaTiO$_3$/SrRuO$_3$
     heterostructures.
\end{abstract}



\maketitle

\section{Introduction}

     The search for alternative gate dielectric materials to replace silica
     (SiO$_2$) in microelectronic devices is one of the grand challenges
     that the materials science community and the Si-based semiconductor
     industry are facing at the current time ~\cite{MRS-bulletin-02}.
     The rapid scaling of the physical gate lengths of
     Metal-Oxide-Semiconductor
     Field-Effect-Transistors (MOSFET) requires a concomitant
     rapid reduction of the gate dielectric thickness
     in order to preserve a high gate oxide capacitance.
     This can no more be accomplished by lowering the size of
     the SiO$_2$ layer because, together with problems in the thickness control,
     the leakage current would become inacceptably high.
     Indeed, the leakage current from the channel to the gate
     is due to the direct tunneling of carriers and increases
     exponentially with the decrease of both
     the thickness of the gate dielectric and
     the height of the electrostatic barrier for
     the electrons through the gate stack.
     The current roadmap projection (assessed by the
     the International Technology Roadmap for Semiconductors, ITRS
     ~\cite{ITRS-01}) imposes the choice of
     an alternative gate dielectric with a good capacitance for a thick-enough
     layer and its full implementation into the production line by 2005.
 
     The properties that the new dielectric should
     meet are well established and have been reviewed recently by Wilk
     and Wallace~\cite{Wilk-01}. They can
     be divided into {\it fundamental material properties}, and
     {\it device processing and performance properties}.
     Amongst the material properties, we can enumerate
     $(i)$ a higher dielectric constant than amorphous silica
     ($\kappa_{\rm SiO_{2}} = 3.9$) in order to increase
     the capacitance  without decreasing the
     thickness, $(ii)$ large band gaps and band offsets with Si to
     prevent tunneling currents,
     $(iii)$ a good thermodynamic stability in contact with the
     Si substrate,
     $(iv)$ a good quality of the interface with the Si channel,
     which means a small number of electrical defects and
     a low midgap interface state density, and
     $(v)$ film morphology avoiding the formation
     of polycristalline films and grain boundaries.
     Amongst the device properties, we can cite
     $(vi)$ a good compatibility with metallic gate electrodes,
     $(vii)$ a compatiblity with the deposition mechanism during the
     fabrication process,
     $(viii)$ reliability.
 
     Many materials satisfy some subset of the
     previous criteria,
     but the identification of a dielectric
     that addresses {\it simultaneously} all of the requirements
     is a real challenge.
     Investigations on
     oxides like  Al$_2$O$_3$, ZrO$_2$, HfO$_2$,
     Ta$_2$O$_5$, Y$_2$O$_3$, Gd$_2$O$_3$, and TiO$_2$,
     have thrown encouraging results in the last
     few years~\cite{Schlom-02}. Amongst the most promising candidates,
     ABO$_3$ perovskite oxides
     (where A stands for Ba or Sr and B stands for Ti)
     appear in good position.

     The ABO$_3$ compounds have a dielectric constant above 300,
     one order of magnitude higher than the other candidates.
     Although they are thermodynamically unstable in direct contact
     with Si (they react to form titanium silicide and alkaline-earth
     silicate~\cite{Schlom-02,Demkov-01}), they can be grown in perfect registry
     with the Si substrate by means of Molecular Beam Epitaxy (MBE)
     when including a silicon-compatible buffer layer.
     On one hand, this layer must be sufficiently thick
     to ensure the physical separation between the substrate
     and the perovskite. On the other hand, it must remain thin enough
     to keep the benefit of the high dielectric constant of the
     ABO$_{3}$ compound (the capacitance of the lower-$\kappa$ buffer
     layer being in series with that of the perovskite).
 
     In the McKee-Walker process~\cite{McKee-98,McKee-01,McKee-patent-98},
     the buffer consists in few atomic layers of AO alkaline-earth
     oxide that can eventually be alloyed during the
     growth in order to accomodate the lattice mismatch with Si.
     The growth of AO on Si includes the presence, at the interface,
     of a (sub)monolayer of ASi$_{2}$ silicide so that the final
     structure corresponds to the sequence Si/ASi$_{2}$/AO/ABO$_{3}$.
     The epitaxy is such that
     $\rm ABO_3 \left( 001 \right) \parallel AO \left( 001 \right)
     \parallel Si \left( 001 \right)$, and
     $\rm ABO_3 \left< 110 \right> \parallel AO \left< 100 \right>
     \parallel Si \left< 100 \right>$, i. e. the ABO$_3$ atomic
     planes are rotated 45$\rm ^o$ around the $\left( 001 \right)$ AO direction
     ~\cite{McKee-91}.
     The epitaxial crystalline growth at the oxide/semiconductor
     interface avoids the formation of defects and ensures
     the continuity of the dielectric displacement ~\cite{McKee-01}.
     MBE techniques allow the control of the growing sequence
     at the submonolayer level preventing grain-boundaries and
     providing a good quality interface and extremly smooth surface morphology.

     First attempts to make MOSFETs including perovskite oxides have
     been reported recently. Using a 110 \AA-thick SrTiO$_3$ layer as the
     gate dielectric, Eisenbeiser {\it et al} ~\cite{Eisenbeiser-00}
     have fabricated a transistor that behaves comparably to a 8
     \AA-thick SiO$_2$/Si MOSFET.
     The improvement in transistor performance was very satisfactory,
     and the leakage currents was two order of magnitudes smaller than
     in a similar SiO$_2$-based device.

     As it was pointed out before, the barrier height of the dielectric
     with respect to the Si substrate should be large enough to minimize
     carrier injection into the conduction band states.
     A large value of the Conduction Band Offset, CBO,  between
     Si and the gate dielectric is required, and typically
     materials with CBO smaller than 1.0 eV are rejected for further
     applications. Robertson and Chen ~\cite{Robertson-99},
     aligning the Charge Neutrality Levels (CNL)~\cite{Tersoff-84} of both
     semiconductors, have estimated
     the CBO for a Si/SrTiO$_3$ interface to -0.14 eV
     (SrTiO$_3$ below, that is no barrier at all for the electrons)
     in very good agreement with experimental results~\cite{Chambers-01}.
     This prevents, in principle, the use of the titanate as the
     gate dielectric in electronic devices.
     However, the presence of the buffer alkaline-earth
     oxide in the heterostructure was missing in their approach.
     In this paper, we will
     show that in addition to providing a physical separation between Si
     and the perovskite, the presence of the alkaline-earth oxide also
     allows to monitor efficiently the band offset.

     We report a study of the
     properties of BaO/BaTiO$_3$ (from now on, we will refer to this
     heterostructure as the
     Ba-interface), and SrO/SrTiO$_3$ (Sr-interface) structures from
     first-principles.
     The method on which the simulations are based is described
     in Section ~\ref{section:technicalities}.
     In Section ~\ref{section:structure}, we discuss
     the details of the atomic structure at the interfaces.
     The electronic structure is presented in Sections
     \ref{section:electronicstructure}, where we analyze the
     density of states at the junctions.
     In Section \ref{section:bandoffset} we study the band-offset
     at the interface. Finally, in Section ~\ref{section:whole},
     an estimate of the band alignment of the whole 
     whole Si/SrO/SrTiO$_3$/Pt and
     Si/BaO/BaTiO$_3$/SrRuO$_3$ structures will be given.

     \begin{table*}[htbp]
       \caption[ ]{Reference configuration and cutoff radii
                  (in bohr) of the pseudopotentials
                  used in our study.
                  Because of the inclusion of the semi-core states
                  in valence, and within the Troullier-Martin scheme,
                  Ba, Ti, and Sr pseudopotentials must be generated
                  for ionic configurations (ionic charge +2).
                  However, these are more suitable than the neutral ones,
                  given the oxidation numbers of these atoms in the
                  alkaline-earth oxides and perovskites.
                    }
       \label{table:pseudo}
       \begin{center}
       \begin{tabular}{|cc|c|c|c|c|}
       \hline
       \hline
            &
            &
        Ba  &
        Sr  &
        Ti  &
         O  \\
         Reference &
                   &
         $5s^{2}, 5p^{6}, 5d^{0}, 4f^{0}$ &
         $4s^{2}, 4p^{6}, 4d^{0}, 4f^{0}$ &
         $3s^{2}, 3p^{6}, 3d^{2}, 4f^{0}$ &
         $2s^{2}, 2p^{4}, 3d^{0}, 4f^{0}$ \\
         \hline
          Core radius (a.u.) &
          s                  &
          1.75               &
          1.50               &
          1.30               &
          1.15               \\
                             &
          p                  &
          2.00               &
          1.50               &
          1.30               &
          1.15               \\
                             &
          d                  &
          2.50               &
          2.00               &
          1.30               &
          1.15               \\
                             &
          f                  &
          2.50               &
          2.00               &
          2.00               &
          1.50               \\
       \hline
       \hline
       \end{tabular}
       \end{center}
     \end{table*}

\section{Technicalities}
\label{section:technicalities}

     Our calculations have been performed within Density Functional
     Theory (DFT) ~\cite{Hohenberg-64} and the Local Density
     Approximation (LDA) ~\cite{Kohn-65}. We used
     a Numerical Atomic Orbital (NAO) method, as it is implemented
     in the {\sc Siesta} code ~\cite{Soler-02,SanchezPortal-97,Ordejon-96}.
     The exchange-correlation functional was approximated using the Perdew
     and Zunger ~\cite{Perdew-81} parametrization of Ceperley-Alder
     data~\cite{Ceperley-81}.

      Core electrons were replaced by {\it ab-initio} norm-conserving
     fully-separable~\cite{Kleinman-82}  Troullier-Martin~\cite{Troullier-91}
     pseudopotentials. Due to the large overlap between
     the semi-core and valence states, the
     $3s$ and $3p$ electrons of Ti, $4s$ and $4p$ electrons
     of Sr, and $5s$ and $5p$ electrons of Ba were explicitly included
     in the calculation. Ti, Sr and Ba pseudopotentials were
     generated scalar-relativistically.
     The reference configuration and cutoff radii
     for all the atoms we used are shown in Table-\ref{table:pseudo}.

       The one-electron Kohn-Sham eigenstates
     were expanded in a basis of strictly-localized ~\cite{Sankey-89}
     Numerical Atomic Orbitals ~\cite{Artacho-99}.
     Basis functions were obtained by finding the eigenfunctions
     of the isolated atoms confined within
     the new soft-confinement spherical potential proposed
     in Ref. ~\onlinecite{Junquera-01}.
     We used
     single-$\zeta$ basis set for the semicore states of Ti, Sr and Ba,
     and double-$\zeta$ plus polarization for the valence states
     of all the atoms. For Sr (respectively Ba) an extra
     shell of $4d$ (respectively $5d$) orbitals was added.
     All the parameters that define the shape and the range
     of the basis functions for Ba, Ti and O were obtained by a
     variational optimization in cubic bulk BaTiO$_3$, following the
     procedure described in Ref. ~\onlinecite{Junquera-01}.
     For Sr, another optimization was performed in bulk SrTiO$_3$,
     frozen in the atomic orbitals of Ti and O to these previously
     optimized in BaTiO$_3$ ~\cite{Note-basis}.

       The electronic density, Hartree and exchange-correlation
     potentials, as well as the corresponding matrix elements
     between the basis orbitals, were
     calculated in an uniform real-space grid
     ~\cite{Soler-02}.
     An equivalent plane wave cutoff of 200 Ry
     was used to represent the charge density.
     Once self-consistency was reached,
     the grid was refined (reducing the distance between
     grid points by half) to compute the total energy,
     atomic forces and stress tensor.

       The integrals in reciprocal space were well converged,
     using in all the cases a sampling in $\vec{k}$
     of the same quality as the $\rm \left( 6 \times 6 \times 6 \right)$
     Monkhorst-Pack ~\cite{Monkhorst-76} mesh in bulk $\rm BaTiO_{3}$.
     The equivalent cutoff-length ~\cite{Moreno-92}
     for this sampling, 13 \AA, was the one employed
     in all simulations.
     This represents a large number of $\vec{k}$-points thought that
     all the materials involved in the heterojunctions
     are insulators. However
     it has been proved that this fineness is mandatory
     while dealing with perovskites ~\cite{King-Smith-94}.

       Test of the performance of the {\sc Siesta} method on perovskites
     were done in bulk BaTiO$_{3}$~\cite{Ordejon-01}. Lattice constants,
     ferroelectric distorsions, Born effective charges,
     and phonon dispersion curves are in very good agreement
     with plane waves ~\cite{King-Smith-94,Ghosez-99.2,Ghosez-95,Zhong-94}
     and full potential LAPW calculations ~\cite{Cohen-90}.

\section{Atomic structure at the interface}
\label{section:structure}

       In Table ~\ref{table:latcon} we report the experimental and
     calculated lattice parameters of
     the different materials involved in our heterostructures,
     together with the lattice mismatch with respect to the 
     Si substrate. The misfit is defined as $f =
     100 \times \left( a - a_{Si}\right) / a_{Si}$,
     where $a$ and $a_{Si}$ are, respectively, the lattice constant
     of the epilayer and Si. The value of
     $f$ is positive when the epilayer is compressed and negative
     when it is expanded.
     In Table~\ref{table:latcon}, we observe that the LDA produces
     a systematic underestimate of the lattice constant
     (about 1$\%$). Nevertheless, the correct sequence of lattice mismatch
     is obtained so that the calculations will reproduce the experimental
     strain conditions when working at the theoretical lattice constants
     of the substrate.

\begin{table}
       \caption[ ]{ Experimental and theoretical lattice constants
     ($a$, in \AA) for the different compounds involved in
     our heterostructures. The lattice mismatch, $f$, between
     a given epilayer and the Si substrate (in \% with respect to the
     substrate lattice constant) is also reported.
     $d_{A-A}$ (A = Ba or Sr),
     stands for A-A nearest neighbour distance in AO oxides.
     Perovskite values refer to the cubic structure.
                  }
\begin{center}
\begin{tabular}{lcccc}
\hline
\hline
       System                           &
       \multicolumn{2}{c}{Experimental} &
       \multicolumn{2}{c}{LDA-DFT}      \\
                                        &
       $a$ (\AA)                   &
       $f$   (\%)                  &
       $a$ (\AA)                   &
       $f$   (\%)                  \\
\hline
       Si                       &
       5.43 ~\footnotemark[1]   &
                                &
       5.389~\footnotemark[2]   &
                                \\
       BaO                      &
       5.52 ~\footnotemark[1]   &
       1.66                     &
       5.433                    &
       0.82                     \\
                                &
       ($d_{Ba-Ba} = 3.90$)     &
                                &
       ($d_{Ba-Ba} = 3.842$)    &
                                \\
       BaTiO$_3$                &
       4.00 ~\footnotemark[3]   &
       4.18                     &
       3.948                    &
       3.60                     \\
\hline
       Si                       &
       5.43 ~\footnotemark[1]   &
                                &
       5.389~\footnotemark[2]   &
                                \\
       SrO                      &
       5.16 ~\footnotemark[1]   &
      -4.97                     &
       5.075                    &
      -5.83                     \\
                                &
       ($d_{Sr-Sr} = 3.65$)     &
                                &
       ($d_{Sr-Sr} = 3.588$)    &
                                \\
       SrTiO$_3$                &
       3.91 ~\footnotemark[4]   &
       1.83                     &
       3.874                    &
       0.90                     \\
\hline
\hline
\end{tabular}
\end{center}
\footnotetext[1]{N. W. Ashcroft, and N. D. Mermin, Ref.~\onlinecite{Ashcroft}.}
\footnotetext[2]{J. M. Soler {\it et al.}, Ref.~\onlinecite{Soler-02}.}
\footnotetext[3]{G. H. Kwei {\it et al.}, Ref.~\onlinecite{Kwei-93}.}
\footnotetext[4]{T. Mitsui {\it et al.}, Ref.~\onlinecite{Landolt-Bornstein}.}
\label{table:latcon}
\end{table}

      Interfaces were simulated using a supercell approximation.
     The basic unit cell, periodically repeated in space
     corresponds to the generic (AO)$_n$/(AO-BO$_2$)$_m$ formula, where
     $n$ and $m$ are respectively the number of AO oxide atomic planes and
     the number of ABO$_3$ unit cells ~\cite{McKee-01}.
     For even $n$ and odd $m$ (the only cases studied in this work),
     this structure possesses two mirror symmetry planes located on
     the central AO and BO$_2$ layers.
 
       We considered pseudomorphic heterojunctions, so that
     the lattice constant parallel to the plane of the interface,
     $a_{\parallel}$, is assumed to remain the same on both sides of
     the structure.
     The choice of $a_{\parallel}$ allows to treat
     implicitly the mechanical effect of the substrate, which is not
     included explicitely in the calculations.

     To establish the notation, we will call the plane parallel
     to the interface the $(x,y)$ plane, whereas the
     perpendicular direction will be referred to as
     the $z$ axis.

       Under the strain conditions imposed by the Si substrate,
     the epitaxial layers will minimize the elastic energy by elongation
     or compression of the lattice constant along $z$, $a_{\perp}$.
     To determine its value, strain relaxations of the bulk
     unit cells of  AO and ABO$_3$ were
     performed under the constraint of fixed $a_{\parallel}$.
     Since the lattice misfit between the substrate
     and the epilayers is
     small enough to remain in the linear regime,
     the different values of $a_{\perp}$ with
     respect to the in-plane lattice constant can be predicted
     from the Macroscopic Theory of Elasticity (MTE),
     and therefore an estimation of the atomic structure of the interface
     can be done. Following the description of Ref. ~\onlinecite{VandeWalle-86},
     and for an interface orientation along (001):

     \begin{eqnarray}
        a_{i,\perp} & = & \left[ 1 - D_{i} 
                    \varepsilon_{i, \parallel} \right] a_i
        \nonumber \\
        \varepsilon_{i, \parallel} & = & \frac{a_{i,\parallel}}{a_{i}} - 1
        \nonumber \\
        D_{i} & = & 2 \frac{c_{12}^{i}}{c_{11}^{i}}
        \label{eq:normallat}
     \end{eqnarray}

      \begin{table}
         \caption[ ]{ Theoretical values of the elastic constants
                    $c_{11}$ and $c_{12}$ in Mbar.
                    }
         \begin{center}
         \begin{tabular}{lccc}
         \hline
         \hline
                              &
         $c_{11}$             &
         $c_{12}$             &
         2($c_{12}/c_{11}$)  \\
         \hline
         BaO       &
          2.10     &
          0.57     &
          0.54     \\
         BaTiO$_3$ &
           3.71    &
           1.26    &
           0.68    \\
         SrO       &
          2.36     &
          0.57     &
          0.48     \\
         SrTiO$_3$ &
           3.93    &
           1.17    &
           0.59    \\
         \hline
         \hline
         \end{tabular}
         \end{center}
         \label{table:elasticconstants}
      \end{table}

     \noindent where $a_{i}$, $c_{11}^{i}$ and $c_{12}^{i}$ stand for,
     respectively, the equilibrium lattice parameter and the elastic constants
     of material $i$.
     Theoretical values of the elastic constants are reported
     in Table ~\ref{table:elasticconstants}. Bulk structures from the
     macroscopic theory are in excellent agreement with the
     first-principles results,
     as can be drawn from the results in Table ~\ref{table:normallat}
     (relative errors within 1 \% for all the cases).

\begin{table}
       \caption[ ]{ Lattice constant perpendicular to the
                    plane of the interface, $a_{\perp}$,
                    at different values of the in-plane lattice constant,
                    $a_{\parallel}$. Results from both, first-principles
                    structural minimizations ($FP$) and
                    macroscopic theory of elasticity ($MTE$)
                    are reported. Units in \AA.
                  }
\begin{center}
\begin{tabular}{lccc}
\hline
\hline
       System                           &
       $a_{\parallel}$                  &
       $a_{\perp}^{MTE}$                &
       $a_{\perp}^{FP}$                 \\
\hline
                                        &
        5.389                           &
        5.457                           &
        5.457                           \\
        BaO                             &
        5.430                           &
        5.435                           &
        5.433                           \\
                                        &
        5.665                           &
        5.307                           &
        5.322                           \\
\hline
                                        &
        3.811                           &
        4.041                           &
        4.054                           \\
        BaTiO$_3$                       &
        3.839                           &
        4.022                           &
        4.025                           \\
                                        &
        4.006                           &
        3.909                           &
        3.911                           \\
\hline
                                        &
        5.389                           &
        4.924                           &
        4.939                           \\
        SrO                             &
        5.430                           &
        4.905                           &
        4.923                           \\
                                        &
        5.522                           &
        4.861                           &
        4.893                           \\
\hline
                                        &
        3.811                           &
        3.912                           &
        3.915                           \\
        SrTiO$_3$                       &
        3.839                           &
        3.895                           &
        3.893                           \\
                                        &
        3.904                           &
        3.856                           &
        3.857                           \\
\hline
\hline
\end{tabular}
\end{center}
\label{table:normallat}
\end{table}

     The resulting bulk tetragonal unit cells were
     used as the building blocks of our supercell.
     However, as interplanar distances in the region
     close to the interface are
     not predicted properly from MTE ~\cite{Peressi-93},
     a full relaxation of the geometry using first principles
     methods was needed.
 
     For each interface, a {\it reference} ionic configuration
     was defined by piling up truncated bulk strained materials.
     Atomic coordinates were then relaxed until the maximum
     component of the force on any atom was smaller than 10 meV/\AA.
     The maximum component of the stress tensor
     along $z$ was smaller than $5 \times 10^{-3}$ eV/\AA$^{3}$
     for the Ba-interface and than $7 \times 10^{-3}$ eV/\AA$^{3}$
     for the Sr-interface. It has been confirmed that additional
     relaxation of $a_{\perp}$ (neglected in this work) does not
     produce any significant change.

     In order to characterize the atomic displacements induced
     by the relaxation, we define
     $\delta_z \left(M_{i}\right)$
     (respectively $\delta_z \left(O_{i}\right)$) as
     the displacement of the cation (respectively oxygen)
     along $z$ at layer $i$, with respect to the initial {\it reference}
     configuration. We introduce
     the displacement of the mean position of each atomic plane
     as $\beta_{i} = \left[ \delta_z \left(M_{i}\right) +
     \delta_z \left(O_{i}\right)\right] / 2$, and the change in the
     interplanar distance between consecutive planes $i$ and $j$
     as $\Delta d_{ij} = \beta_{i} - \beta_{j}$.
     The rumpling parameter of layer $i$ describes the movement of the
     ions with respect to the mean position of each atomic plane and
     corresponds to  $\eta_{i} = \left[ \delta_z \left(M_{i}\right) -
     \delta_z \left(O_{i}\right)\right] / 2 $.
     It is positive when the cation $M_{i}$ is above the oxygen,
     and negative otherwise.
 
\begin{figure*}[htbp]
\begin{center}
\includegraphics[width=13cm] {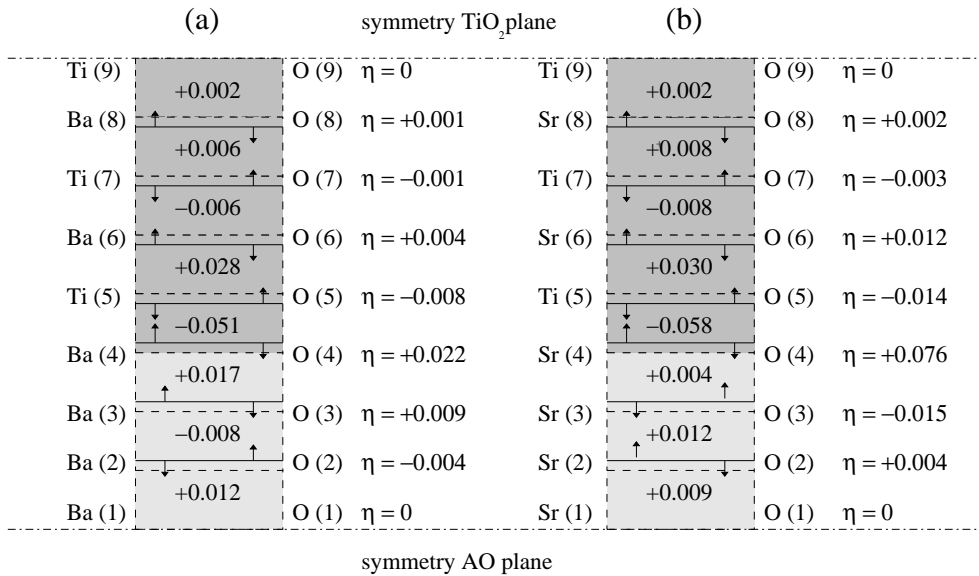}
      \caption{ Schematic view of the atomic relaxation
                for the bottom half of BaO/BaTiO$_3$ (panel a),
                and SrO/SrTiO$_3$ (panel b) supercells.
                Dashed lines correspond to the {\it reference} positions of
                the atomic planes, and the full lines are
                the mean position in the relaxed structure.
                Changes in the interplanar distance are written in \AA.
                The atoms (A or Ti, depending on the layer, at the left and
                O at the right) move in the direction indicated by the arrow.
                The rumpling parameter, $\eta$, is expressed in \AA.
                The size of the heterostructure corresponds to $n$=6, $m$=5.
               }
      \label{fig:relaxations}
\end{center}
\end{figure*}

     Fig.~\ref{fig:relaxations} shows a schematic view of the
     atomic relaxations for both Ba and Sr-interfaces, when
     the in-plane lattice constant was constrained to the
     theoretical one of Si.
     The most important features are:
     $(i)$ a compression of the interplanar distance at the interface layer~;
     $(ii)$ the appearance of an ionic interface dipole, due to the opposite
     motion of the anion and the cation at the AO-layer closest to the
     interface -- the anion moves towards the AO region whereas
     the cations displaces inside the ABO$_3$ part --;
     $(iii)$ a monotonic decay of the absolute value of
     the rumpling parameter as a function of the distance to the
     AO-interface layer, where the major relaxations are localized, and
     $(iv)$ the oscillatory behaviour of the sign of $\eta_{i}$ and
     $\beta_{i}$ inside the perovskite  from layer to layer, as it
     happens also in ABO$_3$ free-standing
     slabs~\cite{Padilla-97,Cheng-00}.

     The main difference between the Ba and the Sr-heterostuctures
     is the magnitude of the relaxations at the interface,
     larger in the last case.
     All these conclusions are independent of the in-plane lattice constant
     imposed in the calculation, and show very good agreement
     with the results obtained using the {\sc Abinit} ~\cite{Abinit} plane-wave
     pseudopotential code. ~\cite{Zimmer-02}.

\section{Electronic structure at the interface}
\label{section:electronicstructure}

\begin{figure}[htbp]
\begin{center}
\includegraphics[width=7cm] {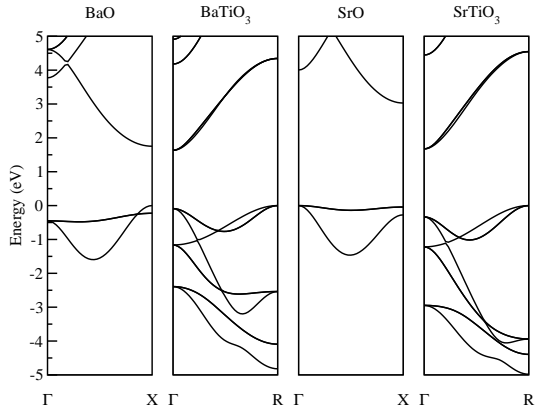}
      \caption{ Bulk band structures of cubic BaO, BaTiO$_3$, SrO, and
                SrTiO$_3$ at the theoretical lattice constant.
                The zero of energy has been asigned
                to the top of the valence band in each case.
                Only the bands closed to the gap are plotted.
               }
      \label{fig:bulkbands}
\end{center}
\end{figure}

      In Figure ~\ref{fig:bulkbands}, we report the energy band structure
    along a selected high symmetry line in the first Brillouin zone
    for the bulk alkaline-earth oxides ($\Gamma$X-line)
    and cubic bulk perovskite structures ($\Gamma$R-line).
    Only bands close to the Fermi level are represented.
    The valence bands are mainly composed of
    O $2p$ states that, in the case of the perovskites,
    show significant hybridizations with Ti $3d$ orbitals.

      All the alkaline-earth and perovskite oxides we consider
     are insulators (experimental gaps above 3 eV).
     Within the cubic symmetry (in the absence of strains)
     and neglecting spin-orbit couplings, the computed band gap for both
     BaTiO$_3$ and SrTiO$_3$
     is indirect ($R \rightarrow \Gamma$).
     The top-most valence and the bottom-most conduction bands
     are three times degenerated at these high symmetry points.
     Under the same conditions,
     the gap is also indirect in SrO, with three degenerated upper most
     valence bands at $\Gamma$ and a single lowest conduction
     band at $X$, whereas BaO exhibits a direct gap,
     between single bands at $X$.

      \begin{table}
         \caption[ ]{ Theoretical ($E_{gap}^{theo}$) and experimental
	             ($E_{gap}^{expt}$) band gaps in eV
                      for the materials involved in our
                      simulations. The theoretical value,
                      within LDA, has been calculated at
                      the theoretical lattice constant.
                    }
         \begin{center}
         \begin{tabular}{ccccc}
         \hline
         \hline
                          &
         BaO              &
         BaTiO$_3$        &
         SrO              &
         SrTiO$_3$        \\
         $E_{gap}^{theo}$ &
         1.75             &
         1.63             &
         3.03             &
         1.67             \\
         $E_{gap}^{expt}$ &
         4.8~\footnotemark[1] &
         3.2~\footnotemark[2] &
         5.7~\footnotemark[1] &
         3.3~\footnotemark[3] \\
         \hline
         \hline
         \end{tabular}
         \end{center}
         \label{table:gaps}
         \footnotetext[1]{W. H. Strehlow, and E. L. Cook,
                          Ref.~\onlinecite{Strehlow-73}.}
         \footnotetext[2]{S. H. Wemple, Ref.~\onlinecite{Wemple-70}.}
         \footnotetext[3]{R. A. McKee, F. J. Walker, and M. F. Chisholm,
                          Ref.~\onlinecite{McKee-01}.}
      \end{table}

     In Table ~\ref{table:gaps} we report experimental
     and theoretical bands gaps (within LDA) for all the materials
     involved in our study. We  see that,
     due to the well known DFT ``band gap problem'',
     the theoretical values are understimated by about 50 \%
     in each case. Nevertheless, it is usually accepted that this
     error can be roughly compensated by an appropriate shift of the
     conduction bands which should not
     affect the conclusions of the character of the gap reported in this
     Section.

     A uniaxial strain along (001) lowers the symmetry of the perovskites from
     Pm3m to P4mm.
     This translates into a splitting of the top of the valence
     bands into a singlet and a doublet.
     The singlet is above (below) for a compressive (tensile) strain.
     For the alkaline-earth oxides, the symmetry reduces from
     Fm$\overline{3}$m 
     to I4/mmm. The top of the valence band of SrO is therefore
     split but, in this case, the doublet is
     above (below) the singlet
     for a compressive (tensile) strain.
     Spin-orbit couplings (not considered in this work)
     might introduce further splittings.

\begin{figure*}[htbp]
\begin{center}
\includegraphics[width=13cm] {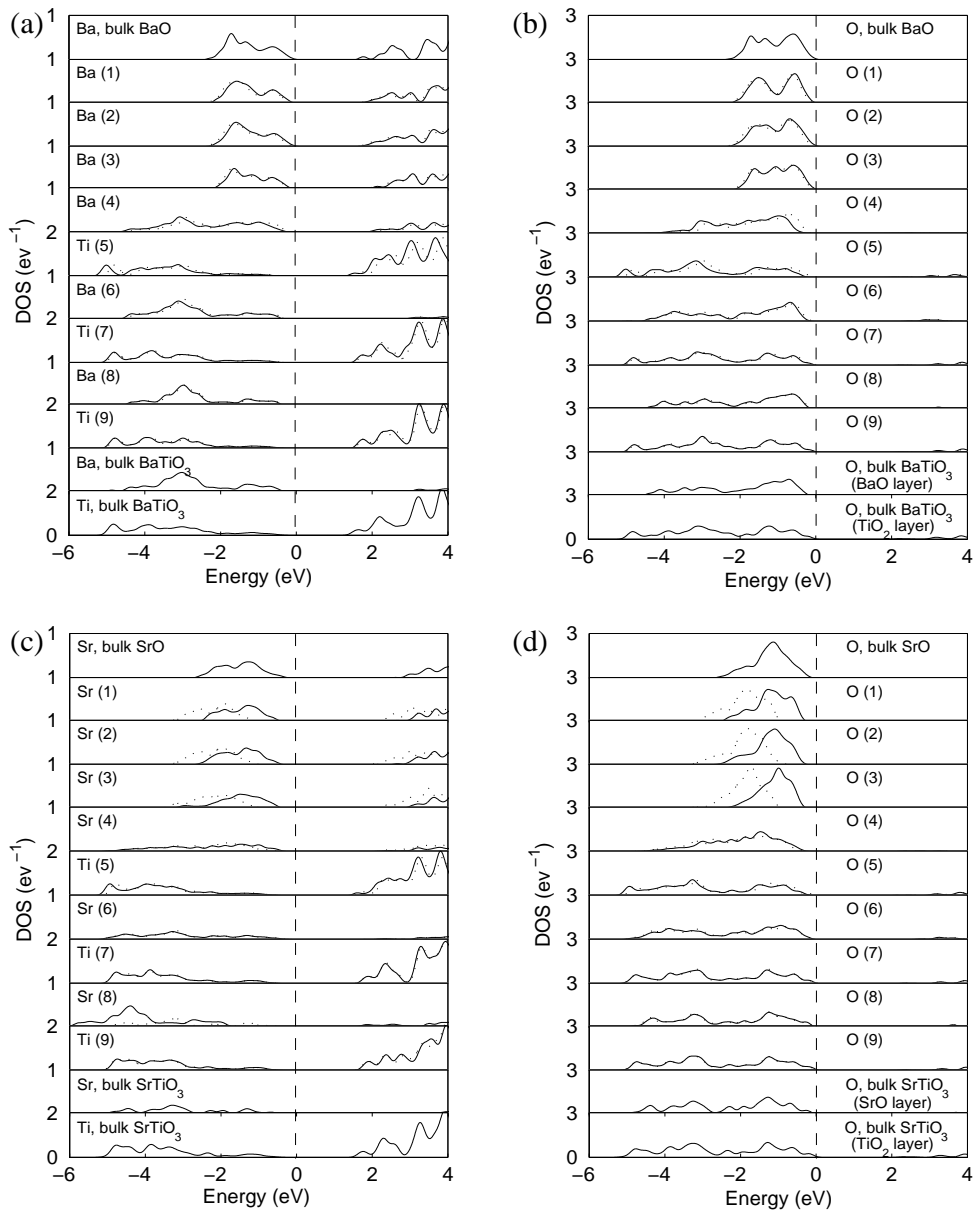}
      \caption{ Projected Density Of States
                on all the atoms as a function of the distance
                to the interface for the BaO/BaTiO$_3$ (panels a and b)
                and SrO/SrTiO$_3$ (panels c and d) heterostructures.
                Full lines represent the projection for the relaxed geometry and
                dotted lines for unrelaxed coordinates.
                Atomic layers are identified as in Fig. ~\ref{fig:relaxations}.
                Projected Density Of States of bulk AO and ABO$_3$
                are also plotted for comparison.
                All the energies have been rigidly displaced
                in order to align the top of the valence band
                (vertical line) with zero.
                The imposed $a_{\parallel}$ was set up to the theoretical
                one of Si (5.389 \AA).
               }
      \label{fig:dos}
\end{center}
\end{figure*}

     Fig. ~\ref{fig:dos} shows
     the Projected Density Of States (PDOS)
     on the different atoms (sum of the projections of the DOS
     on all the atomic orbitals of the given atom) as a function of
     the depth of the layer inside the material for the
     BaO/BaTiO$_3$ and SrO/SrTiO$_3$ interfaces.
     The main conclusions that can be extracted are as follows:
     $(i)$ the absence of any interface induced gap states
     clearly demonstrates the semiconductor character of the heterostructures;
     $(ii)$ the features of the PDOS on the alkaline-earth and the O atom
     at the interface layer (labelled as 4) are much closer to the ones
     displayed in bulk-ABO$_3$ than in bulk-AO, showing a dominant
     ABO$_3$ character of the interface;
     $(iii)$ the PDOS converges very quickly to the bulk properties and
     many of the bulk features can be recovered even at the atomic
     layers closest to the interface;
     $(iv)$ atomic relaxations have small effects on the shape
     of the PDOS, as
     can be seen comparing the solid and dotted curves in the figure. 
     Only a shift in the SrO layers towards the zero energy (chosen
     as the top of the valence band in each case)
     is noticeable. 
     This effect is a direct consequence of the relaxation-induced
     interface dipole discussed in Section ~\ref{section:structure}.
     The different magnitude of the dipole between the
     Ba and Sr-interfaces explains why the shift is
     almost negligeable in the Ba-heterostructure.

\section{Band offset}
\label{section:bandoffset}

      One of the most important physical quantities that characterize
     the interface between semiconductors or insulators is the band offset,
     i. e., the relative position of the energy levels on both sides
     of the interface. The valence-band offset, VBO (respectively
     conduction-band offset, CBO) is defined as the difference between
     the positions of the top of the valence bands (respectively the bottom
     of the conduction bands) of the two materials.
     These band discontinuities play a fundamental
     role in calculating the transport properties
     through heterojunction devices.

     The determination of these offsets from first-principles cannot be
     achieved from a direct comparison of the corresponding band edges in
     the two compounds as obtained from two independent
     bulk band-structure calculations. The reason is the lack of an
     intrinsic energy scale to refer all the energies: in a
     first-principles simulation, the hamiltonian eigenvalues
     are referred to an average of the electrostatic potential
     that is ill-defined for infinite systems~\cite{Kleinman-81}
     (it is only defined to within an arbitrary constant).
     Consequently, together with the eigenvalue difference,
     we must consider the lineup of this average
     between the two materials. This potential shift 
     depends on the dipole induced by the electronic charge 
     transferred from one part of the 
     interface to the other after the interfacial hybridization
     (the electronic charge density of each system will decay into 
     the other in an, in principle, unknown way). 
     The transfer of charge depends not only on the
     materials that constitute the interface, but also on the 
     particular orientation, so the lineup 
     can only be obtained from a self-consistent
     calculation on a supercell including both materials.

     Therefore, from the theoretical point of view, the band offsets (BO)
     are usually split into two terms ~\cite{Colombo-91,Peressi-98}:

     \begin{eqnarray}
        {\rm BO}  & = & \Delta E_{v,c} + \Delta V
     \end{eqnarray}
 
     The first contribution,
     $\Delta E_{v}$ (resp. $\Delta E_{c}$), is referred to as the
     {\it band-structure term}. It is defined as the difference between
     the top (resp. bottom) of the valence (resp. conduction) bands
     as obtained from two
     independent standard bulk band-structure calculations
     at the same strained geometries as in the supercell construction.
     Within LDA, only a first estimate of the band-structure term can
     be obtained, $\Delta E_{v,c}^{LDA}$.
     To get more accurate results, a correction dealing with many-body
     effects in the quasiparticle spectra should be added:
 
     \begin{equation}
          \Delta E_{v,c} = \Delta E_{v,c}^{LDA} + \Delta E_{v,c}^{corr}
     \end{equation}

     Self-energies corrections are often obtained within the GW
     approximation~\cite{Hybertsen-86}. They strongly modify
     the description of the conduction bands, and tend to solve the
     ``band gap problem'' mentioned in Section
     ~\ref{section:electronicstructure}. Even the valence band energies
     might be subject of certain errors, specially in oxides~\cite{Kralik-98}.
     Unfortunately, no accurate GW data are currently available for AO
     and ABO$_{3}$ compounds. Only model GW calculations have
     been performed recently for SrO and SrTiO$_{3}$ and with 
     limited success~\cite{Capellini-00}.
     To overcome the problem, we make the approximation that the errors
     in the valence bands are smaller than those for the conduction bands
     and of the same order of magnitude for the two compounds taking part
     in the heterostructures so that they tend to cancel each
     other ($\Delta E_{v}^{corr} = 0$). Knowing the relative position of
     the valence bands, we simply add the experimental band gaps
     (see Table ~\ref{table:gaps}) to obtain the discontinuities for
     the conduction bands
     ($\Delta E_{c} = \Delta E_{v}^{LDA} + \Delta E_{gap}^{expt}$).
 
     The second term, $\Delta V$, is the {\it lineup of the
     average of the electrostatic potential} through the
     heterojunction. This macroscopic quantity summarizes all the
     intrinsic interface effects, such as the chemical composition,
     structural details and orientation.
     To obtain it, we start from the total (ionic plus electronic) 
     microscopic electrostatic Hartree
     potential, output of the
     self-consistent supercell calculation
     (in this Section we will define the zero-energy level as 
     the average of this potential in the unit cell ~\cite{deltav}). Then, we
     apply the double-macroscopic
     average~\cite{Baldereschi-88,Colombo-91,Peressi-98} technique.
     It consists of performing first the average of the
     electrostatic potential over planes parallel to the interface,
     and then averaging the obtained quasiperiodic one-dimensional
     function with two step-like filter functions 
     whose lenghts, $l_1$ and $l_2$, are determined by the
     periodicity of the constituents.
     Here $l_1$ and $l_2$ have been set up to the distance between
     equivalent AO and TiO$_2$ planes in the alkaline-earth oxide and
     in the perovskite respectively. 
     A full description of the
     method to the AO/ABO$_3$ heterostructures can be found in
     Ref. ~\onlinecite{Junquera-01.2}.
     The resulting profile of the macroscopic potential is
     flat on both sides far enough from
     the interface ({\it bulk-like regions}).
     $\Delta V$ is defined as the difference between these
     two plateau values (see Fig. \ref{fig:bandoff}).
     The lineup should be independent of the length used in the 
     filter functions. 
     However we have checked how doubling the size of the
     step-like functions introduce a numerical uncertainty in $\Delta V$
     of the order of 30 meV. This is the main source of inaccuracy
     in our calculations of the band offsets. 
%

     It is worth noticing that
     neither $\Delta E_{v}$ nor $\Delta V$ have any physical meaning
     by their own, being pseudopotential dependent numbers.
     Only the sum of both is physically significant and quite independent of
     the choice of the pseudopotential ~\cite{Peressi-98}.

      \subsection{First principles results}

\begin{figure*}[htbp]
\begin{center}
\includegraphics[width=17cm,angle=0] {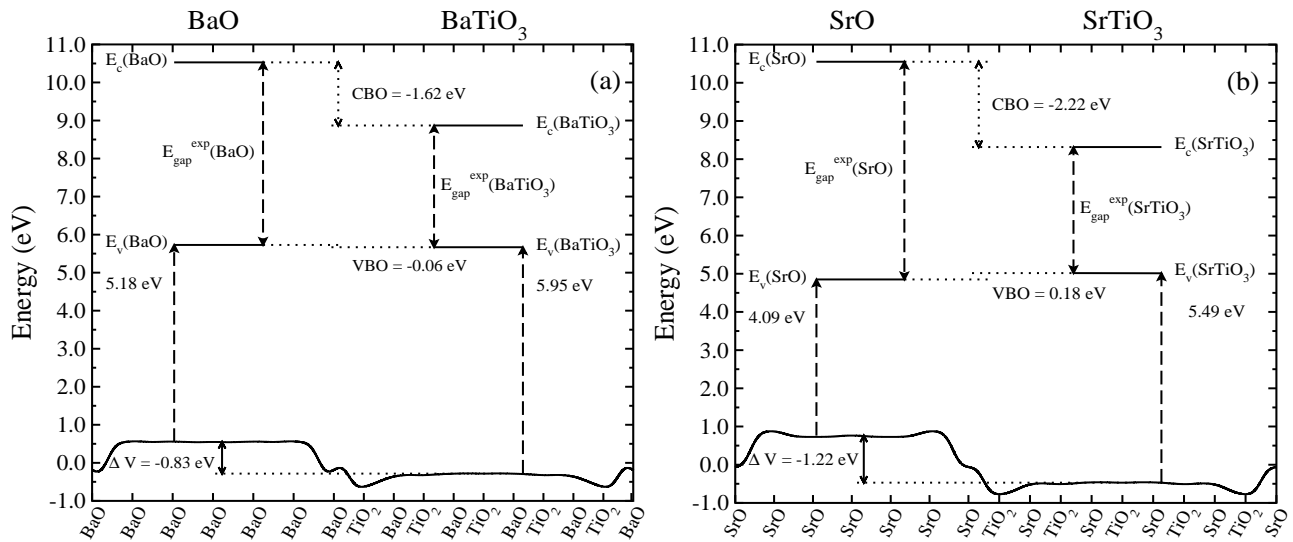}
      \caption{Schematic representation of the valence-band offset (VBO)
               and the conduction-band offset (CBO) for BaO/BaTiO$_3$
               (panel a), and SrO/SrTiO$_3$ interface (panel b).
               E$_{v}$, E$_{c}$, and E$_{gap}^{exp}$ stand
               for the top of the valence band, the bottom of the conduction
               band and the experimental band gap respectively.
               Values for E$_{v}$, measured with respect to the average of the
               electrostatic potential in each material, are indicated.
               The solid curve represents the profile of the macroscopic
               average of the total electrostatic potential
               across the interface. $\Delta V$ stands for the resulting
               lineup.
               The in-plane lattice constant was set up to the 
               theoretical one of
               Si (5.389 \AA). The size of the supercell corresponds
               to $n$ = 6 and $m$ = 5.
               }
      \label{fig:bandoff}
\end{center}
\end{figure*}

     Fig. ~\ref{fig:bandoff} shows a schematic
     representation of the band structure discontinuities
     for the BaO/BaTiO$_3$ and SrO/SrTiO$_3$ heterostructures,
     both of them calculated fixing the in-plane lattice constant
     at the theoretical one of Si.
     Band splittings steming from strain
     are taken into account in the figure.
     We use the same sign convention as Van de Walle and Martin in
     Ref. ~\onlinecite{Vandewalle-87}: a positive value of the band offset
     for the discontinuity at a junction A/B corresponds to an upward
     step in going from A to B.

        From the figure, we conclude that Ba-interface
     is type-II, with both the valence and conduction
     bands of BaTiO$_3$ falling in energy below the corresponding ones of BaO.
     Within the accuracy of our calculations, BaO and BaTiO$_3$ topmost
     valence bands are almost aligned (an offset of only -0.06 eV is predicted),
     so the barrier in the conduction bands is mainly due to the difference
     in the band gaps of both materials and, inferred from the experimental
     values, amounts to -1.62 eV.
 
        Sr-interface is type-I,
     meaning that the band gap of SrTiO$_3$
     lies completely inside the gap of SrO.
     An upward step of +0.18 eV for the valence bands is theoretically expected,
     which implies a CBO of -2.22 eV.

       A rough estimate of the valence band offset
     was already accessible ~\cite{Rao-97} by
     identifying in Fig. ~\ref{fig:dos}
     the position of the top of the valence band
     in the PDOS for the O atom at both symmetry planes,
     in the bulk-like regions
     of the materials that constitute the interface (numbered as
     1 for AO and 9 for ABO$_3$). The values deduced from
     the figure are -0.09 eV for
     the Ba-interface (BaO above) and +0.28 eV for the Sr-interface (SrTiO$_3$
     above), close to those obtained using the
     macroscopic average technique.
     However these numbers must be taken with care~\cite{Peressi-98}:
     this method to compute band
     offsets requires calculations with a higher number of special 
     $\vec{k}$-points
     than those needed to converge the charge density or the
     potential lineup.

        It is important to note here the {\it crucial} role played by
     the atomic relaxations at these polar interfaces.
     As was pointed out in section ~\ref{section:structure},
     after the relaxation process an extra dipole appears at the
     junction that modifies the electrostatic lineup accross the
     interface ~\cite{Martin-81} and, consequently, the band offsets~:

     \begin{eqnarray}
	\delta \left( \Delta V \right) & = & \frac{ 4 \pi}{a_{\parallel}^{2}}
        \sum_{\kappa \alpha}
	\frac{Z^{*(T)}_{\kappa,\alpha z} } {\epsilon_{\infty}} 
        \Delta u_{\kappa \alpha}
     \end{eqnarray}

     \noindent where $\delta \left( \Delta V \right)$
     is the change in the electrostatic lineup along $z$-direction due
     to the atomic displacements,
     $Z^{*(T)}_{\kappa, \alpha \beta}$ is the Born effective tensor of atom
     $\kappa$, $\Delta u_{\kappa \alpha}$ its displacement along 
     cartesian direction $\alpha$ during the
     relaxation  and $\epsilon_{\infty}$ the optical dielectric
     constant. Looking at the magnitude of the atomic
     displacements, it is reasonable that the change should be more
     remarkable for the Sr-interface than for the Ba one.
     From our {\it ab-initio} calculations, and for the same supercell
     used to get results in Fig. ~\ref{fig:bandoff},
     we observe  a change in $\Delta V$ of -0.67 eV for the Sr-interface
     (from 1.16 eV for the unrelaxed geometry to 0.49 eV after the relaxation),
     whereas in the Ba-interface the deviation amounts to -0.11 eV
     (from 0.44 eV to 0.33 eV).
     This emphasizes the importance of performing accurate first-principles
     atomic relaxations for correct predictions of the barriers.

\begin{table}
       \caption[ ]{ Valence-band offsets (VBO) for BaO/BaTiO$_3$
                    and SrO/SrTiO$_3$ interfaces. Values are reported
		   at different in-plane lattice constants, $a_{\parallel}$.
                    $\Delta E_{v}$ and $\Delta V$ stand for, respectively,  the
                    band structure term and the line up of the electrostatic
                    potential contributions to VBO.
                    The size of the heterostructures corresponds
                    to $n$ = 6, $m$ = 5.
                  }
\begin{center}
\begin{tabular}{lcccc|ccc}
\hline
\hline
                                  &
\multicolumn{4}{c|}{BaO/BaTiO$_3$} &
\multicolumn{3}{c}{SrO/SrTiO$_3$} \\
\hline
$a_{\parallel}$ (\AA) &
5.389                &
5.430                &
5.583                &
5.665                &
5.389                &
5.430                &
5.522                \\
\hline
$\Delta V$           &
-0.834               &
-0.833               &
-0.746               &
-0.755               & 
-1.217               &
-1.190               &
-1.128               \\
$\Delta E_{v}$       &
  0.772               &
  0.807               &
  0.600               &
  0.560               & 
  1.401               &
  1.340               &
  1.209               \\
VBO                  &
-0.062               &
-0.026               &
-0.146               &
-0.195               & 
0.184                &
0.150                &
0.081                \\
\hline
\hline
\end{tabular}
\end{center}
\label{table:bandofflatcon2}
\end{table}

\begin{figure}[t]
\begin{center}
\includegraphics[width=7cm] {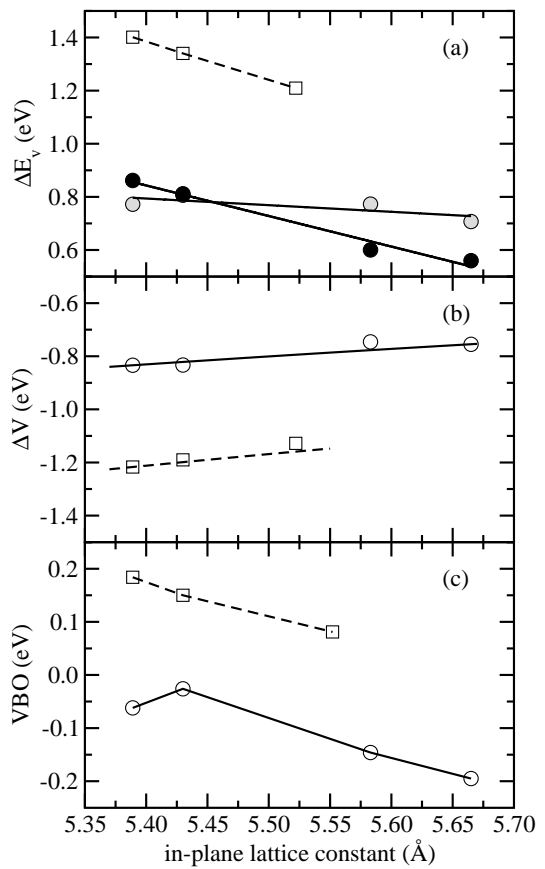}
   \caption{ Dependence with in-plane strain of the valence-band offset (VBO),
             and of its decomposition into the band-structure ($\Delta
             E_v$), and lineup ($\Delta V$) terms.
             Squares and circles represent, respectively,
             the first-principles results
             for the SrO/SrTiO$_3$ and BaO/BaTiO$_3$ interfaces.
             In panel (a), grey-filled (respectively black-filled) 
             circles stand for the
             diffrence between the top of the valence band of BaTiO$_3$
             and the highest occupied state at X (respectively Z)
             point in BaO.
             Lines in panels (a) and (c) (dashed for Sr and full for
             the Ba-interface), are a guide to the eye.
             Lines in panel (b) represent the results
             of the anayltic scaling law proposed 
             in Ref. ~\onlinecite{DiVentra-96.2}.
            }
   \label{fig:strain}
\end{center}
\end{figure}

     To what extent do these discontinuities change with the in-plane lattice
     constant? This is an important question because a dependence
     with strain would allow us
     to tune the band offsets (for example, replacing the Si
     substrate by Ge~\cite{McKee-01} in order
     to impose a different lattice parameter throughout the interface)
     depending on the required values for
     a given device.
     In order to check this point,
     we have carried out calculations at
     different in-plane lattice constants.
     In Table ~\ref{table:bandofflatcon2} and Fig. ~\ref{fig:strain}
     we summarize the results for both, Ba and Sr-heterostructures.
     In both cases a variation by about 0.1 eV in VBO
     with the in-plane lattice constant is observed,
     mainly due to the band-structure term
     (consequence of the strain-induced splittings of the
     top valence-band manifold), as it happens for
     other lattice-mismatched, isovalent, common anion interfaces
     ~\cite{DiVentra-96.1}.
     The change is almost linear, and
     tends to lower the energy of the valence bands of the ABO$_3$
     perovskite with respect to the AO alkaline-earth oxide.

     The band-structure term displays a linear behaviour with strain
     for the Sr-interface (see Fig. ~\ref{fig:strain}-panel a).
     The anomalous behaviour of $\Delta E_v$ for the Ba-interface
     is due to a modification in the character of the top
     of the valence band of BaO under strain.
     It changes from $X$ when BaO is compressed to $Z$ when
     it is expanded. This transformation occurs for
     a lattice constant around 5.43 $\rm \AA$ (theoretical lattice
     parameter of BaO).
     In Fig. ~\ref{fig:strain}(a) we plot the difference between the
     top of the valence band of BaTiO$_3$ and the highest occupied
     state at $X$ and $Z$ of BaO. The crossing point is
     clearly identified in the figure.
     No extra changes in the linear beaviour of $\Delta E_v$ are
     expected for longer lattice constants.
 
     The almost-linear change in the lineup term can
     be explained according to an analytic scaling law
     proposed in Ref. ~\onlinecite{DiVentra-96.2}.
     Once $\Delta V$ is known for a reference configuration with
     an in-plane lattice constant $a_{\parallel}$,
     then, supposing an uniform strain throughout the structure,
     $\Delta V'$ for any other strained
     configuration $a^{'}_{\parallel}$ can be extrapolated from:

\begin{equation}
    \Delta V^{'} \simeq \frac{1}{\left( 1 +
                 \varepsilon_{\parallel}^{2} \right)}
                 \Delta V
    \label{eq:scalinglaw}
\end{equation}

     \noindent where $\varepsilon_{\parallel}
     = (a^{'}_{\parallel}/a_{\parallel} - 1)$.
     Fig. ~\ref{fig:strain}(b) shows a comparison of the
     first-principles and extrapolated values, where
     the heterostructure at the in-plane lattice constant of
     Si has been chosen as the reference configuration.
     Results are in good agreement within the numerical
     accuracy of the {\it ab-initio} results.

     In summary, the VBO varies almost linearly for a large range
     of in-plane  strains. The only deviation is observed for
     BaO/BaTiO$_{3}$ and is explained by a change of character
     of the BaO gap under compression.
 
     \section{Interface with Si}
     \label{section:whole}

     As it was pointed out in the Introduction,
     AO/ABO$_3$ interface is only a part of the
     gate stack of technological interest for the semiconductor industry.
     AO acts as a buffer layer between the Si substrate and
     the high-$\kappa$ perovskite.
     The whole heterostructure epitaxially grown
     following the McKee-Walker process ~\cite{McKee-patent-98}
     is made of Si/ASi$_2$/AO/ABO$_3$
     ~\cite{McKee-98}. As it will be emphasized in this Section,
     the role of the buffer layer is not only the
     passivation of the Si substrate, but also the efficient tuning
     of the offsets between the perovskite and the channel.
 

\begin{figure*}[htbp]
\begin{center}
\includegraphics[width=15cm] {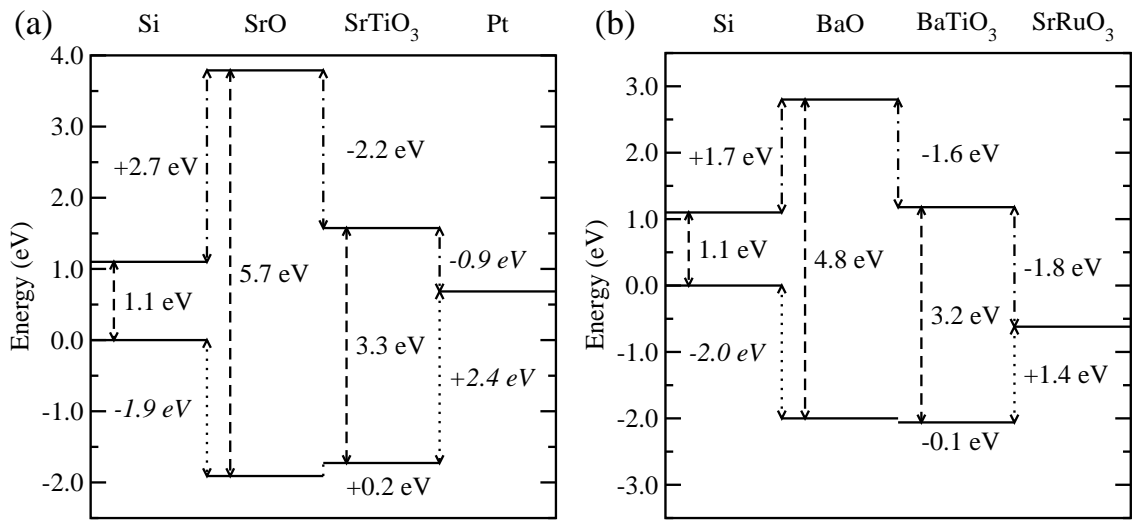}
      \caption{ Estimation of the valence (dotted lines)
                and conduction (dot-dashed lines) band offsets
                for the whole heterostructures
                Si/SrO/SrTiO$_3$/Pt (panel a), and
                Si/BaO/BaTiO$_3$/SrRuO$_3$ (panel b).
                Dashed lines represent the experimental band gaps.
                Theoretical value for the VBO between Si and AO
                (in italic) has been taken from
                Ref. ~\onlinecite{Nardelli-02} for the Sr interface, and from
                Ref. ~\onlinecite{Gulleri-02}
                for Ba-interface.
                Theoretical Schottky-barriers between SrTiO$_3$ and Pt
                (also in italic) 
                have been taken from Ref. ~\onlinecite{Robertson-99}.
               }
      \label{fig:wholehet}
\end{center}
\end{figure*}

     Combining our results with various data available in the
     literature, we can estimate the band discontinuities along the
     whole heterostructures of technological interest as summarized
     in Fig. ~\ref{fig:wholehet}.
     Previous theoretical calculations of the band offsets
     between the alkaline-earth oxide AO and Si have been reported
     recently (Si/BaO~\cite{Gulleri-02}, Si/SrO~\cite{Nardelli-02}.
     In addition, we can find in the literature theoretical
     estimations for the Schottky barriers
     between perovskites and prototypical metallic electrodes
     (SrTiO$_3$/Pt ~\cite{Robertson-99}, or
     BaTiO$_3$/SrRuO$_3$~\cite{Junquera-02}).

     We must notice that, altough most of
     the previous works (except Ref. ~\onlinecite{Robertson-99})
     have been done using the same basic approaches
     (DFT, pseudopotentials, $\vec{k}$-point samplings,
     supercells, etc),
     they differ in the details of the calculations (for
     example, they have been carried out at different in-plane lattice
     constant, and the size of the supercells, or the parameters
     used to generate the pseudopotentials
     might change from one to the other).
     So, only a rough estimate of the barriers can be deduced from the
     comparison and any quantitative conclusion
     is beyond the scope of this Section.
 
     Within LDA, plus GW corrections,
     Boungiorno-Nardelli {\it et al.} ~\cite{Nardelli-02} investigated
     the structural and electronic
     properties of the Si/SrSi$_2$/SrO interface.
     They predicted a VBO between Si and SrO of
     -1.91 eV for the most stable interface configuration.
     Using the experimental gaps to locate the conduction bands,
     it translates in a CBO of 2.69 eV.
     The Schottky barrier $\phi_n$ (difference between the Fermi level and
     the bottom of the conduction band)
     between SrTiO$_3$ and Pt has been evaluated
     ~\cite{Robertson-99} to -0.89 eV, which implies a barrier
     height $\phi_p$ (difference between the Fermi level and the
     top of the valence band) of 2.41 eV. These results are summarized
     in the first panel of Figure~\ref{fig:wholehet}.
 
     Through first-principles gradient-density-functional calculations
     Gulleri {\it et al.} ~\cite{Gulleri-02} focused on the
     structure, energetic and band offsets of the Si/BaO interface.
     For the favoured geometry, they obtained a VBO of -2.0 eV.
     Some of us evaluated the Schottky barriers between BaTiO$_3$
     and SrRuO$_3$ (a typical metallic electrode in
     ferroelectric devices ~\cite{Eom-93b}) to be equal to
     $\phi_p$ = +1.44 eV and $\phi_{n}$ = -1.76 eV. These results are
     summarized in the second panel of Figure~\ref{fig:wholehet}.

     For both stacks, we can clearly see how the problem of the large
     carrier injection (expected for the perovskite in direct contact
     with Si~\cite{Robertson-99}) is overcome by the use of the AO layer.
     The electrostatic barriers for both the electrons and holes,
     between the gate electrode and the channel are
     large enough to prevent carrier injections and
     to push the use of ABO$_3$ perovskites to a prominent position
     to replace silica as the gate dielectric oxide in MOSFETs.

     \section{Conclusions}

     We have studied structural and electronic properties of
     BaO/BaTiO$_3$ and SrO/SrTiO$_3$ interfaces
     from first-principles.
     Atomic relaxations have been performed.
     Interface dipoles, due to the opposite motion of the
     anion and cation atoms at the interface,
     appear for both heterostructures.
     No interface electronic states
     are induced in the band gap.
     The character of the AO layer at the interface is mainly
     perovskite-like.
     Under the experimental strain conditions, the valence bands
     of BaO and BaTiO$_3$ are almost aligned
     (within the accuracy of our calculations),
     whereas a slightly larger barrier is predicted for SrO/SrTiO$_3$.
     Interface dipoles, induced by atomic relaxations,
     have a strong effect on the band alignments at the interface.
     A nearly linear variation of the VBO with in-plane strain is
     observed.
 
     Gathering together our results and various data available in the
     literature, we make a guess for the band alignment of
     whole Si/SrO/SrTiO$_3$/Pt and Si/BaO/BaTiO$_3$/SrRuO$_3$ structures.
     In both cases large enough electrostatic barriers
     for electrons and holes between the gate electrode
     and the channel are estimated, preventing the injection of carriers
     and suggesting that both perovskites compounds are promising candidates
     to replace silica in MOSFETs. 
     Our results should be confirmed by more accurate calculations
     for the whole heterostructure.


     \section{Acknowledgments}
     The authors are indebted with
     Maria Peressi, Luciano Colombo, Rodney McKee, Marco Buongiorno-Nardelli,
     Hermann Kohlstedt, Darrel Schlom, Alex Demkov, and Jun Wang for
     useful discussions. Motorola PSRL supported the 
     initial stages of this work. 
     This work was supported by the Volkswagen-Stiftung
     (www.volkswagenstiftung.de) within the program ``Complex Materials:
     Cooperative Projects of the Natural, Engineering, and Biosciences" with the
     title: ``Nano-sized ferroelectric Hybrids" under project number I/77 737.
     Three of us (J. J., M. Z., and Ph. G.)
     acknowledge support from FNRS-Belgium (grants 9.4539.00 and 2.4562.03)
     and the Universit\'e  de Li\`ege (Impulsion grant).
     J. J. and P. O. acknowledge finantial support from
     the Fundaci\'on Ram\'on Areces and Spanish MCyT grant BFM2000-1312.
     Calculations have been performed on the NIC computer at the
     Universit\'e  de Li\`ege, and on computational resources
     from CESCA and CEPBA.


\begin{thebibliography}{63}
\bibitem[{Not()}]{Note-basis}
\bibinfo{note}{Values for the parameters can be obtained upon request.}
\bibitem[{del()}]{deltav}
\bibinfo{note}{As it is usual within first-principles codes, in this paper we
  define the zero-energy level as the $\vec{G} = \vec{0}$ Fourier-component (i.
  e., the mean value over the unit cell) of the total (ionic plus electronic)
  electrostatic potential, $V^H \left( \vec{r} \right)$. This is not the case
  within the {\sc Siesta} method, where the reference is assigned to the
  average of $\delta V^H \left( \vec{r} \right)$, the electrostatic potential
  generated by the deformation charge density, $\delta \rho \left( \vec{r}
  \right)$, defined ~\cite{Soler-02} as the difference between the
  self-consistent electronic charge density and the sum of atomic densities,
  $\rho ^{atom} = \sum_I \rho ^{atom}_{I}$. The difference between $V^H \left(
  \vec{r} \right)$ and $\delta V^H \left( \vec{r} \right)$ (the so-called the
  Neutral Atom potential in Ref. ~\onlinecite{Soler-02}) is a short-range local
  component, (therefore a bulk quantity not dependent in the details of the
  interface). In this work, to remain compatible with our choice for the
  reference, its lineup is included in the band-structure term.}
\end{thebibliography}
\end{document}